\documentclass[aps,twocolumn,prd,showpacs,showkeys,preprintnumbers,superscriptaddress,nobibnotes,nofootinbib,floatfix,longbibliography]{revtex4-1}

\usepackage{graphicx}
\usepackage{bm}
\usepackage{times}
\usepackage{hyperref}
\usepackage{slashed}
\usepackage{color}
\usepackage{aas_macros}
\usepackage{nicefrac}

\usepackage{slashed}
\usepackage{lipsum}
\usepackage{subfigure}
\usepackage{multirow}
\usepackage{amsmath}
\usepackage{array}    
\usepackage{varwidth} 
\usepackage{comment}

\hypersetup{
    pdfnewwindow=true,
    colorlinks=true,
    linkcolor=blue,
    citecolor=blue,
    filecolor=blue,
    urlcolor=blue
}
\newcommand{\be}{\begin{equation}}
\newcommand{\ee}{\end{equation}}

\newcommand{\eVdist}{\kern-0.06em}
\newcommand{\mev}{\:\text{Me\eVdist V}}
\newcommand{\gev}{\:\text{Ge\eVdist V}}
\newcommand{\dbar}{\bar{\text{d}}}
\newcommand{\hebar}{\overline{\text{He}}}
\newcommand{\hebarthree}{\textsuperscript{3}\overline{\text{He}}}
\newcommand{\hebarfour}{\textsuperscript{4}\overline{\text{He}}}

\begin{document}

\title{Dark Matter Annihilation Can Produce a Detectable Antihelium Flux through $\bar{\Lambda}_b$ Decays}
\author{Martin Wolfgang Winkler}
\email{martin.winkler@su.se, ORCID: orcid.org/0000-0002-4436-0820}
\affiliation{Stockholm University and The Oskar Klein Centre for Cosmoparticle Physics,  Alba Nova, 10691 Stockholm, Sweden}
\author{Tim Linden}
\email{linden@fysik.su.se, ORCID: orcid.org/0000-0001-9888-0971}
\affiliation{Stockholm University and The Oskar Klein Centre for Cosmoparticle Physics,  Alba Nova, 10691 Stockholm, Sweden}

\begin{abstract}
\noindent Recent observations by the Alpha Magnetic Spectrometer (AMS-02) have tentatively detected a handful of cosmic-ray antihelium events. Such events have long been considered as smoking-gun evidence for new physics, because astrophysical antihelium production is expected to be negligible. However, the dark-matter-induced antihelium flux is also expected to fall below current sensitivities, particularly in light of existing antiproton constraints. Here, we demonstrate that a previously neglected standard model process --- the production of antihelium through the displaced-vertex decay of $\bar{\Lambda}_b$-baryons --- can significantly boost the dark matter induced antihelium flux. This process can triple the standard prompt-production of antihelium, and more importantly, entirely dominate the production of the high-energy antihelium nuclei reported by AMS-02.
\end{abstract}

\maketitle

\section{Introduction}

The detection of massive cosmic-ray antinuclei has long been considered a holy grail in searches for WIMP dark matter~\cite{Donato:1999gy, vonDoetinchem:2020vbj}. Primary cosmic-rays from astrophysical sources are matter-dominated, accelerated by nearby supernova, pulsars, and other extreme objects. The secondary cosmic-rays produced by the hadronic interactions of primary cosmic-rays can include an antinuclei component, but the flux is highly suppressed by baryon number conservation and kinematic constraints~\cite{Duperray:2005si, Gomez-Coral:2018yuk}. Dark matter annihilation, on the other hand, occurs within the rest frame of the Milky Way and produces equal baryon and antibaryon fluxes~\cite{Donato:1999gy, Baer:2005tw, Brauninger:2009pe, Cui:2010ud}

The preeminent target for cosmic-ray antinucleon searches is antideuterium, which inhabits a ``goldilocks zone" where the specificity of the dark matter signal is high and the flux is potentially detectable~\cite{Baer:2005tw}. Significant satellite-~\cite{Battiston:2008zza} and balloon-based~\cite{Fuke:2005it, Aramaki:2015laa} efforts have been undertaken to potentially detect and distinguish antideuterons from background events. More recently, $\hebarthree$ has also been considered as a potential dark matter detection target, though even optimistic models based on shower-averaged nucleon statistics indicate that the $\hebarthree$ flux should fall several orders of magnitude below that of $\dbar$~\cite{Carlson:2014ssa, Cirelli:2014qia}.

Recently, however, the Alpha Magnetic Spectrometer (AMS-02) has announced the potential detection of $\mathcal{O}(10)$ $\hebarthree$ events, along with $\sim$2 potential $\hebarfour$ events~\cite{AMSLaPalma}. Such a detection could revolutionize the search for dark matter, as the probability of such a signal from astrophysical mechanisms is vanishingly small~\cite{Chardonnet:1997dv,Duperray:2005si,Cirelli:2014qia,Herms:2016vop,AMSLaPalma, Blum:2017qnn,Korsmeier:2017xzj,Poulin:2018wzu}. Unfortunately, however, dark matter models are naively expected to also produce $\ll$1~detectable antihelium event~\cite{Carlson:2014ssa, Cirelli:2014qia}. Because the verification of a detectable antihelium flux would qualify as bulletproof evidence of new physics, several studies have investigated adaptations to boost the antihelium rate, including changes to the coalescence momentum of $\hebarthree$~\cite{Coogan:2017pwt}, the effects of cosmic-ray reacceleration~\cite{Cholis:2020twh}, extended dark sectors~\cite{Heeck:2019ego}, or the consideration of exotic sources like primordial antimatter clouds~\cite{Poulin:2018wzu}.

In this \emph{letter}, we challenge the current understanding that standard dark matter annihilation models cannot produce a measurable antihelium flux. Our analysis examines a known, and potentially dominant, antinuclei production mode which has been neglected by previous literature -- the production of antihelium through the off-vertex decays of the $\bar{\Lambda}_{b}$. Such bottom baryons are generically produced in dark matter annihilation channels involving $b$ quarks. Their decays efficiently produce heavy antinuclei due to their antibaryon number and 5.6~GeV rest-mass, which effectively decays to multi-nucleon states with small relative momenta. Intriguingly, because any $\hebarthree$ produced by $\bar{\Lambda}_b$ inherits its boost factor, these nuclei can obtain the large center-of-mass momenta necessary to fit AMS-02 data~\cite{AMSLaPalma}.

\vspace{0.25cm}

\emph{State of the Field --} We first examine why such decays have been neglected in previous studies. Dark matter annihilation is usually assumed to proceed to quark or gluon states. Their subsequent hadronization produces a cascade of (meta)stable final-state particles including antiprotons and antineutrons. Rarely, these antiparticles may fuse to form heavier antinuclei. Because the dynamics of antinuclei formation are complex, a ``coalescence model" is typically utilized, where antiparticles formed with relative momenta below a coalescence momentum $p_c$ are assumed to combine to form heavier antinuclei. The value of $p_c$ is typically tuned to collider data.

Due to the small production rate of antideuterons and (especially) antihelium nuclei, most analyses calculate antinuclei spectra using an ``event-averaged" approach. These studies first calculate the flux of individual antiprotons and antineutrons, and subsequently cross-match these particles between events to calculate the antinuclei yield. This technique produces high-precision spectral measurements, but assumes that individual antiparticle momenta are uncorrelated -- an assumption that is strongly violated by off-vertex decays.

Several studies have also investigated coalescence in an event-by-event framework, often with the intention of verifying the accuracy of the event-mixing technique~\cite{Carlson:2014ssa, Cirelli:2014qia,Kachelriess:2020uoh}. However, these analyses have exclusively focused on prompt antinucleus production at the initial vertex. Antinuclei originating from long-lived intermediate resonances have been rejected to prevent cross-mixing between particles produced with similar momenta but different phase-spaces. As a consequence, the decay of single intermediate particles into multi-nucleon final states has been ignored.

\noindent \emph{Methodology -- } In this study, we examine the antihelium flux produced by dark matter annihilation on an event-by-event basis. We improve upon previous techniques by including contributions from multiple antiparticle states produced at single displaced vertices.

To derive the antihelium spectrum, we employ two state-of-the-art event generators: Pythia (version 8.2) and Herwig (version 7.2). Herwig offers the option to carry out $\Lambda_b$ decays with the specialized tool EvtGen, which we also consider.\footnote{In the default configuration, Herwig uses EvtGen for $B$ decays but its own cluster hadronization algorithm for $\Lambda_b$ decays.} While Herwig reproduces the LEP measurement of the transition ratio $f(b\rightarrow\Lambda_b)=0.1^{+0.04}_{-0.03}$~\cite{Caso:1998tx,Abbiendi:1998ip} within $1\sigma$, Pythia falls short by a factor $\sim 3$. Thus, we also consider a tuned version of Pythia (denoted ``$\Lambda_b$-tune") where we increase diquark formation in hadronization ({\tt probQQtoQ}) to match the $\Lambda_b$ production rate at LEP.

Nucleus formation is implemented via an event-by-event coalescence model. The latter contains the coalescence momentum $p_c$ as a single free parameter that determines the phase space window in which coalescence takes place. While $p_c$ is well-defined in the case of antideuterons, which include only two particles, several potential definitions of $p_c$ exist for multi-particle states like antihelium. The authors of Ref.~\cite{Cirelli:2014qia} require that all three particle pairs must have a momentum difference $|p_i-p_j| < p_c$, while Ref.~\cite{Carlson:2014ssa} instead assumes that all three particles must lie within a sphere centered on the center of momentum with a radius $p_c$/2. These results produce antihelium fluxes that differ by $\sim$15\%, which can be accounted for by normalizing the value of $p_c$ to ALICE data.

In this work, we assume antihelium nuclei merge if all particles lie within a momentum sphere of radius 2$^{1/6}p_c$/2. The extra factor of $2^{1/6}$, compared to~\cite{Carlson:2014ssa}, is required to match the definition of $p_c$ in the analytic coalescence model (see~\eqref{eq:BA}).
This assumption is discussed in detail in Appendix~\ref{app:coalescence}.
In addition to the coalescence condition, we require antinucleons to either originate from the initial vertex or from the same parent particle vertex (most importantly the $\bar{\Lambda}_b$).

We determine $p_c$ separately for each Pythia and Herwig implementation via a fit to antideuteron data from ALEPH and antihelium data from ALICE as described in Appendix~\ref{app:coalescence}:
\begin{align}\label{eq:pcpythiaherwig}
  p_c &= 239^{+25}_{-30}\mev\quad \text{(Pythia)}\nonumber\\
  p_c &= 124^{+13}_{-16}\mev\quad \text{(Pythia $\Lambda_b$-tune)}\nonumber\\
  p_c &= 215^{+25}_{-30}\mev\quad \text{(Herwig, Herwig+EvtGen)}
\end{align}
Notice that the diquark parameter in Pythia also significantly boosts prompt antinucleus production.  This is compensated through a reduction of the coalescence momentum by a factor $\sim 0.6$ in the $\Lambda_b$-tune compared to default Pythia.

\begin{figure}[ht!]
\centering
\includegraphics[width=.44\textwidth]{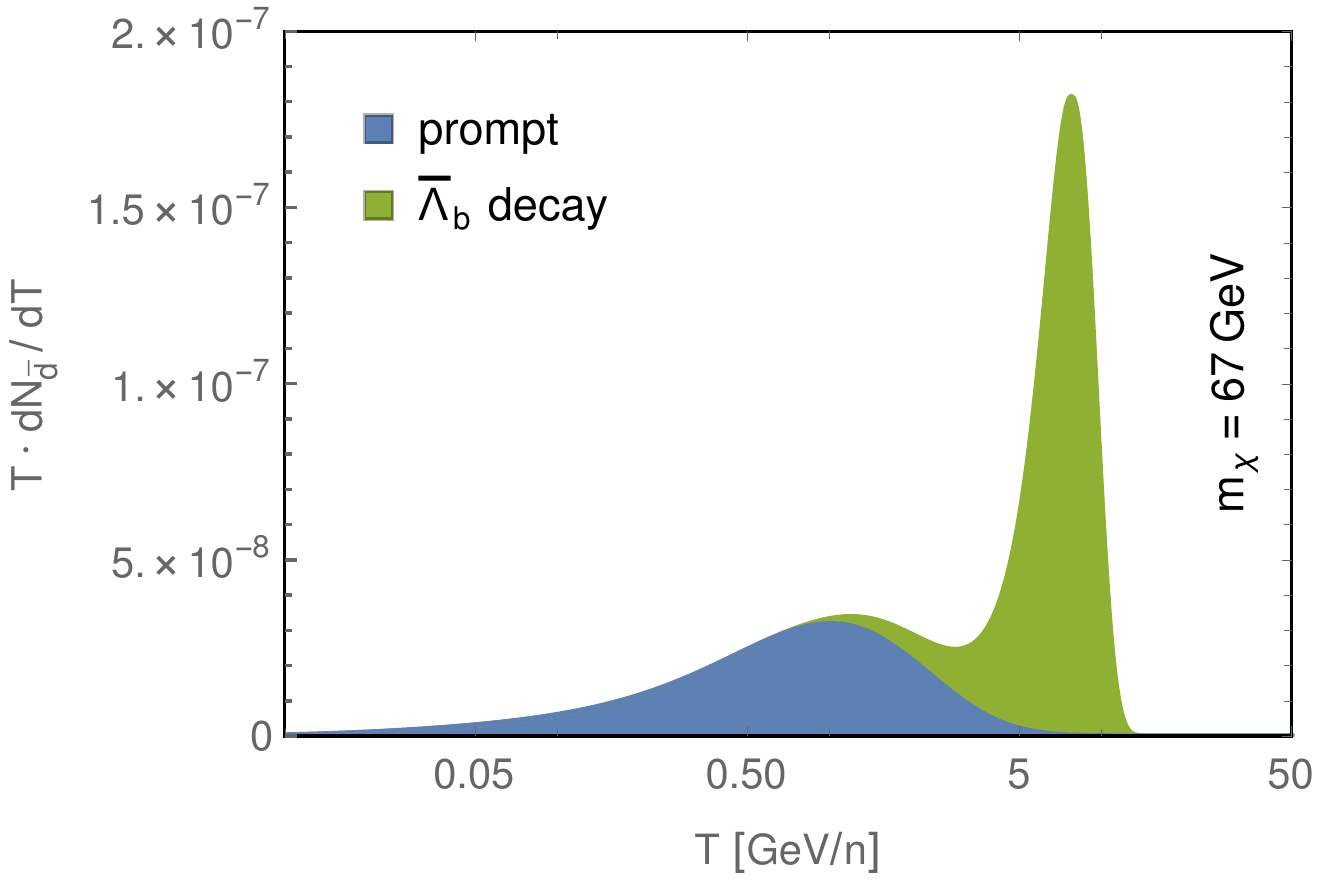}
\vspace{-0.2cm}
\caption{The antihelium injection spectrum from dark matter annihilation using the Pythia event generator. The prompt and $\bar{\Lambda}_b$-induced contributions are shown separately. The antihelium flux from $\bar{\Lambda}_b$ decays exceeds that of prompt events by nearly a factor of two, and entirely dominates the production of high-energy antihelium.}
\label{fig:spectra}
\end{figure}

We stress that the results of our analysis are independent of dark matter mass. However, the expected spectrum of antihelium events will depend on the boost-factor imparted to the $\bar{\Lambda}_{b}$, which depends on the dark matter mass and final annihilation state. As a benchmark scenario, we take $m_\chi= 67\gev$, $\langle \sigma v \rangle=2\times 10^{-26}\:\text{cm}^3/s$ and assume that dark matter annihilates into bottom quarks. These choices are motivated by observations of the Galactic center excess and antiproton excess in similar regions of the dark matter parameter space~\cite{Daylan:2014rsa, Cui:2016ppb, Cuoco:2016eej}. Additionally, we examine a tuned-scenario, where a dark matter particle of initial mass 80~GeV annihilates through light-mediators ($\phi$), with masses of 14~GeV, which subsequently decay to $b\bar{b}$ final states. Such a scenario places a significant fraction of the particle energy directly above the $\bar{\Lambda}_{b}$ mass, increasing the relative antihelium production rate. While we refrain from a detailed fit to AMS-02 antiproton data, we note that our coalescence modeling does not affect the antiproton flux, and our models predict similar antiproton yields as previous work fitting the AMS-02 excess~\cite{Cui:2016ppb, Cuoco:2016eej, Heisig:2020nse}.

\begin{figure*}[ht!]
\centering
\includegraphics[width=.48\textwidth]{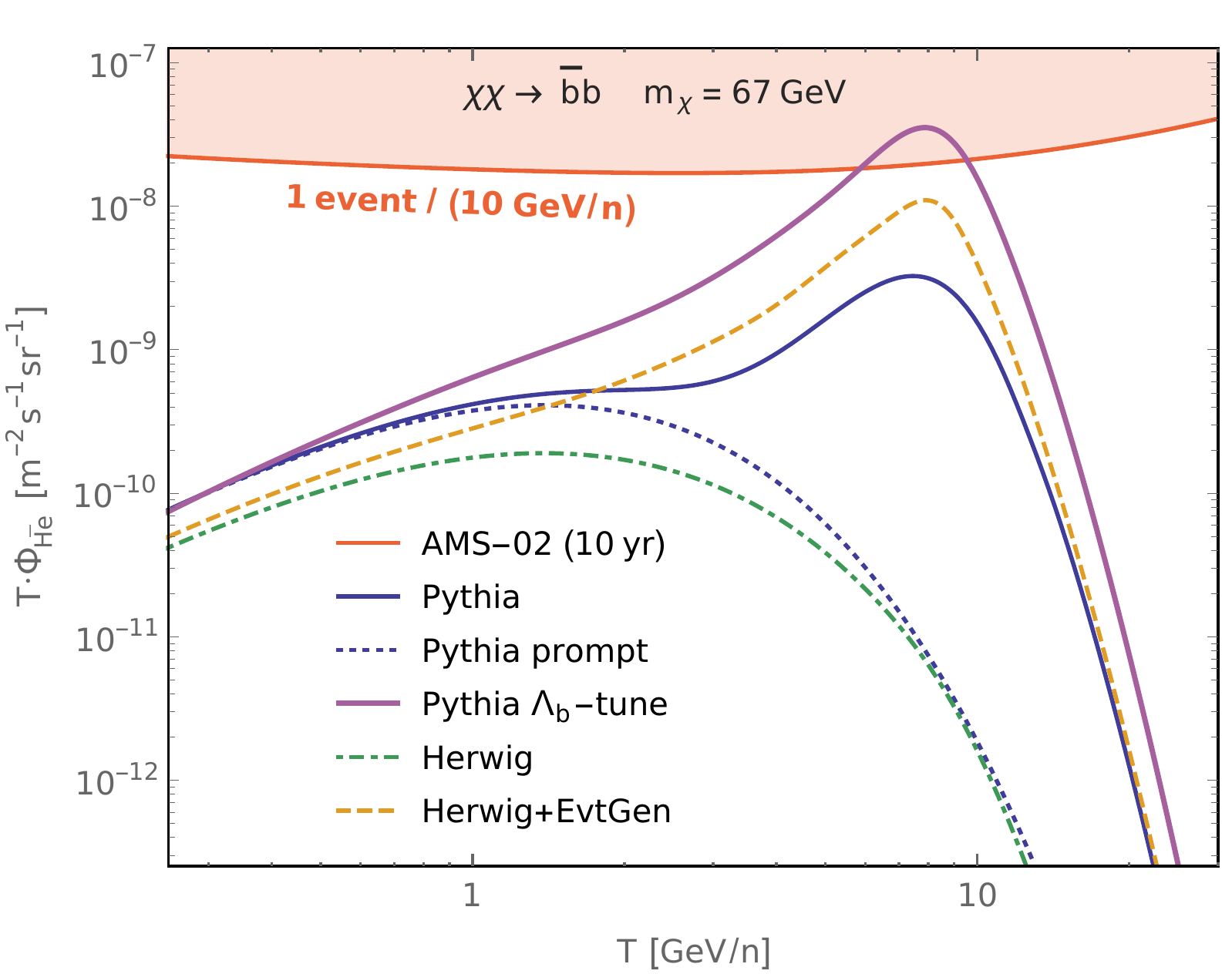}\hspace{0.02\textwidth}
\includegraphics[width=.48\textwidth]{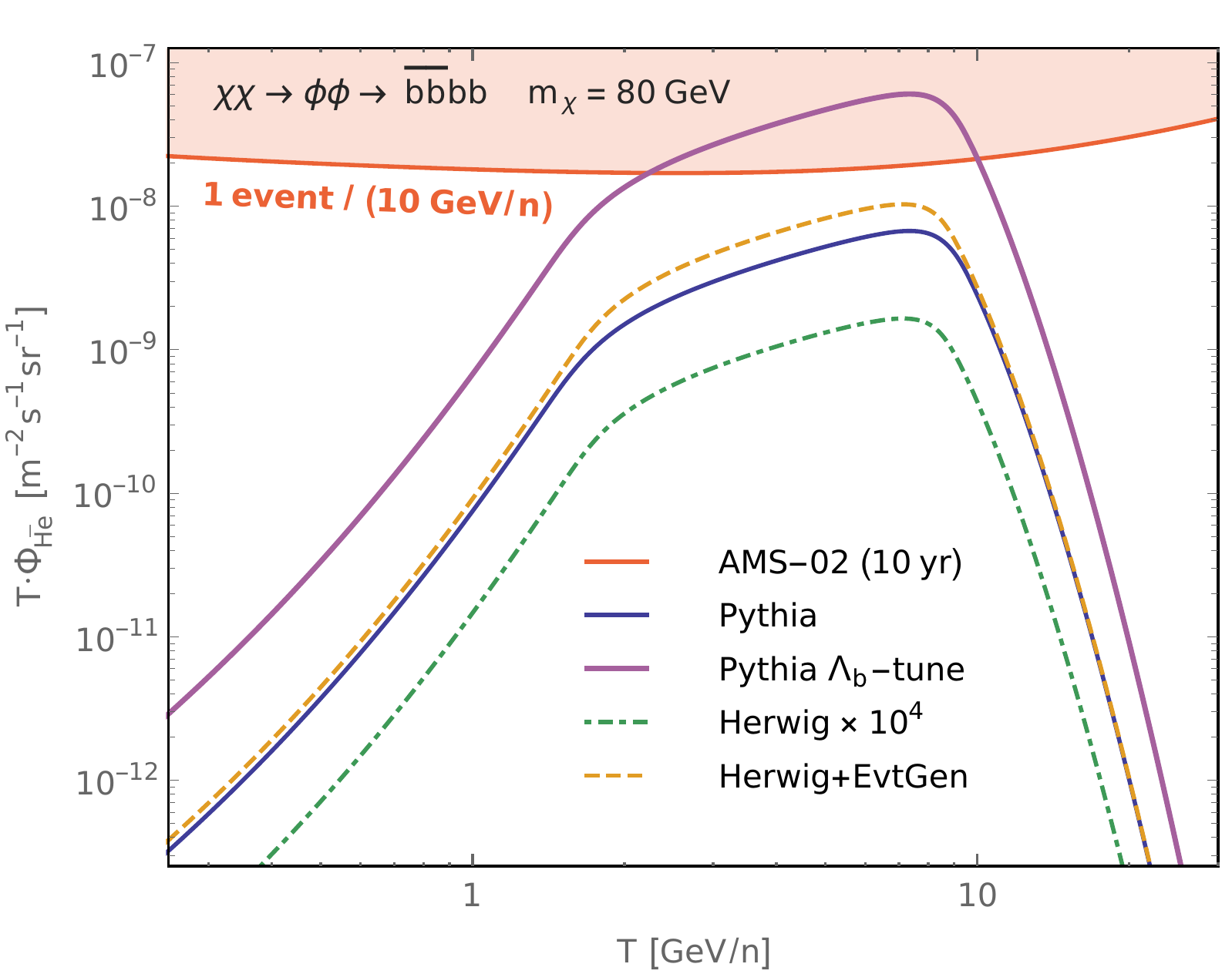}
\vspace{-0.15cm}
\caption{(Left:) The antihelium flux from dark matter annihilation with $\langle \sigma v \rangle=2\times 10^{-26}\:\text{cm}^3/s$, for different choices of event generator as described in the text. The AMS-02 10-year sensitivity taken from~\cite{Kounine2011} is shown as the orange shaded region (see Appendix~\ref{app:hesensitivity}). These results show the significant boost in the number of detectable antihelium events
due to $\Lambda_b$ decays. (Right:) Same, but for dark matter which annihilates through 14~GeV light mediators. For this model, Pythia does not predict any prompt events (all events are produced by $\Lambda_b$ decay).}
\label{fig:fluxes}
\end{figure*}

To calculate the antihelium flux and spectrum at Earth, we must also account for the propagation of antihelium nuclei through the Galactic halo. Because this propagation is not the focus of our study, we employ a standard two-zone diffusion model fit to \mbox{AMS-02} B/C and antiproton data~\cite{Reinert:2017aga,Heisig:2020nse} (see Table~3 in~\cite{Heisig:2020nse}). We normalize the dark matter flux using an NFW density profile~\cite{Navarro:1996gj} with a local density of 0.38~GeV~cm$^{-3}${\bf }~\cite{Catena:2009mf, Pato:2015dua}. The local propagation of low-energy cosmic-rays is severely affected by the heliospheric environment. We account for this effect using an improved force-field description~\cite{Cholis:2015gna} that includes charge-sign effects related to the polarity of the Sun. We note that tritons decay to antihelium before propagating to Earth, and treat this contribution additively to determine the local antihelium flux.

\vspace{0.2cm}

\noindent \emph{Results --} In Figure~\ref{fig:spectra}, we plot the antihelium injection spectrum produced by 67 GeV dark matter annihilation using Pythia. Intriguingly, we find that displaced-vertex $\bar{\Lambda}_b$ decays produce the majority ($\sim$60\%) of the total antihelium flux. More importantly, the boost factor obtained by the $\bar{\Lambda}_b$ is efficiently transferred to antihelium nuclei, dominating the high-energy regime where AMS-02 events are observed.

\begin{table}[hb!]
\begin{center}
\begin{tabular}{|c|c|c|c|c|}
\hline
  &&&&\\[-3.2mm]
  $\quad$Generator$\quad$ & $\quad$P$\quad$  & $\,$P [$\Lambda_b$-tune$]\,$ & $\quad$H$\quad$&$\quad$H+EvtGen$\quad$\\[0.5mm]
\hline &&&&\\[-3.2mm]
 $\hebarthree$ events & 0.1 (0.007) & 0.9 & 0.003 & 0.3 \\[0.5mm]
\hline &&&&\\[-3.2mm]
  $\dbar$ events & 3.7 (3.5)& 4.2 & 1.7 & 2.1\\[0.5mm]
 \hline
\end{tabular}
\end{center}
\vspace{-0.4cm}
\caption{The expected number of $\hebarthree$ and $\dbar$ events from dark matter annihilation with 10~yr of AMS-02 data, for our four choices of event generator. For default Pythia (P), we also list prompt events in brackets. While both instances of Pythia and the Herwig+EvtGen model produce a significant enhancement to the antihelium flux, default Herwig (H) models predict a smaller contribution.}
\label{tab:events}
\end{table}

In Table~\ref{tab:events} and Figure~\ref{fig:fluxes} we compare the integrated and differential antihelium flux at Earth with the AMS-02 sensitivity, finding that, in most models, the addition of $\bar{\Lambda}_b$ decays significantly increases the total event rate and predicts the detection of $\mathcal{O}(1)$ antihelium event in current AMS-02 data. Interestingly, while the total antihelium injection rate increases by a factor of a few, the number of detectable antihelium events increases by more than an order of magnitude due to the increased sensitivity of AMS-02 and the decreasing effect of solar modulation at higher energies.

In simulations using the Pythia $\Lambda_b$-tune, as well as our Herwig+EvtGen model, the predicted number of detected antihelium events increases beyond a factor of $100$. However, our default Herwig analysis does not efficiently produce antihelium events through the $\bar{\Lambda}_b$ channel. At first inspection, this is surprising because Herwig produces a $\bar{\Lambda}_b$ flux that is 4x higher than the default Pythia implementation. However, while Br($\bar{\Lambda}_b\rightarrow \hebar)\simeq 3\times 10^{-6}$ in Pythia, this branching ratio resides below $10^{-9}$ in Herwig.

The significant difference between Pythia and Herwig is rooted in the underlying hadronization models. Pythia uses an implementation based on the Lund String Model~\cite{Andersson:1983ia}, while Herwig uses a cluster model from Ref.~\cite{Webber:1983if}. In both Pythia and Herwig, the off-shell weak decay:
\be
\bar{\Lambda}_b \rightarrow \bar{d}\,u\,\bar{u}\,(\overline{ud})_0
\ee
emerges as the dominant channel for displaced antihelium production, where $(\overline{ud})_0$ denotes an antidiquark state. We first note that the branching fraction to this final state is 5$\times$ smaller in Herwig than in Pythia. However, this explains only a small fraction of the difference between these results.

\begin{figure*}[ht!]
\centering
\includegraphics[width=.48\textwidth]{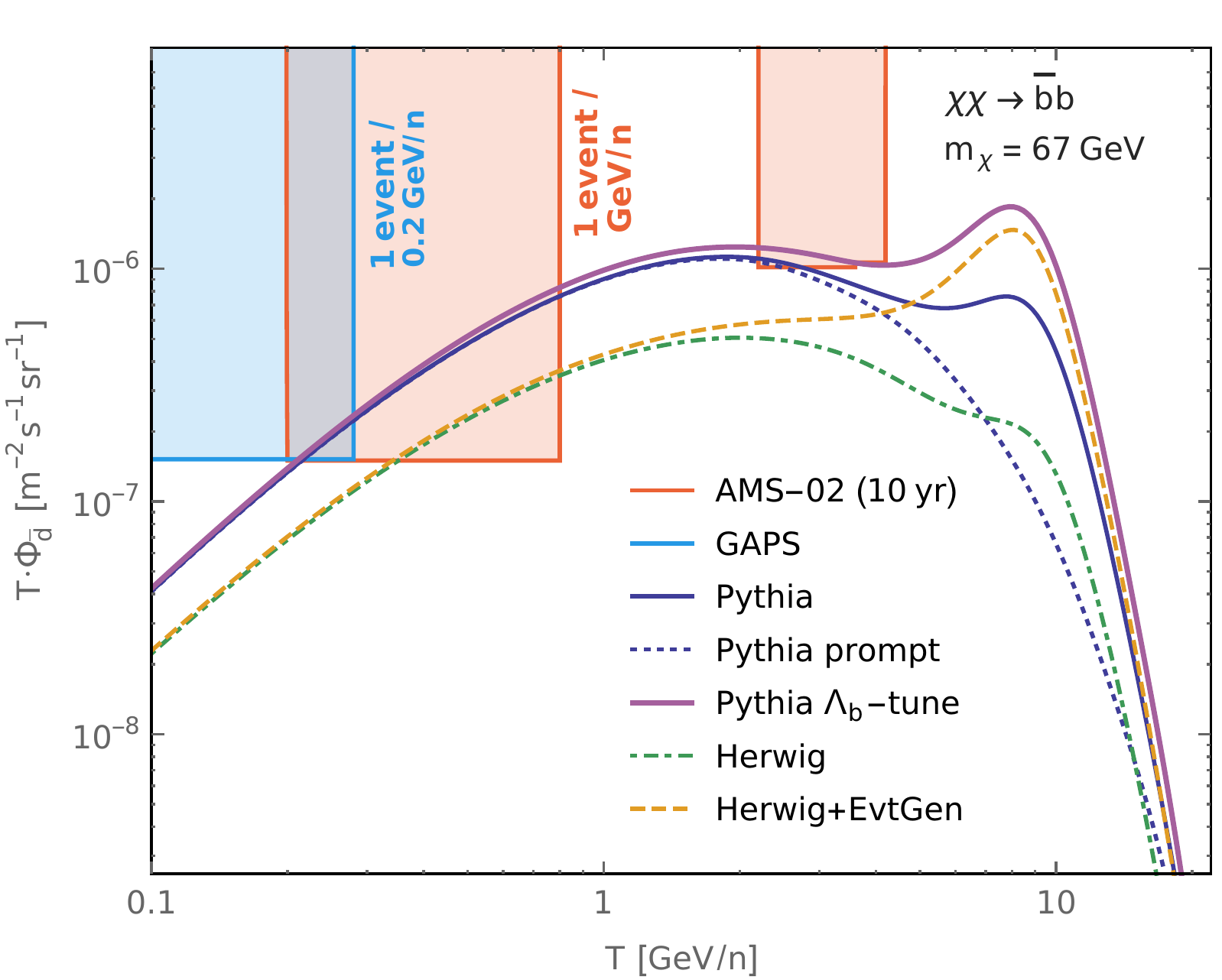}\hspace{0.02\textwidth}
\includegraphics[width=.48\textwidth]{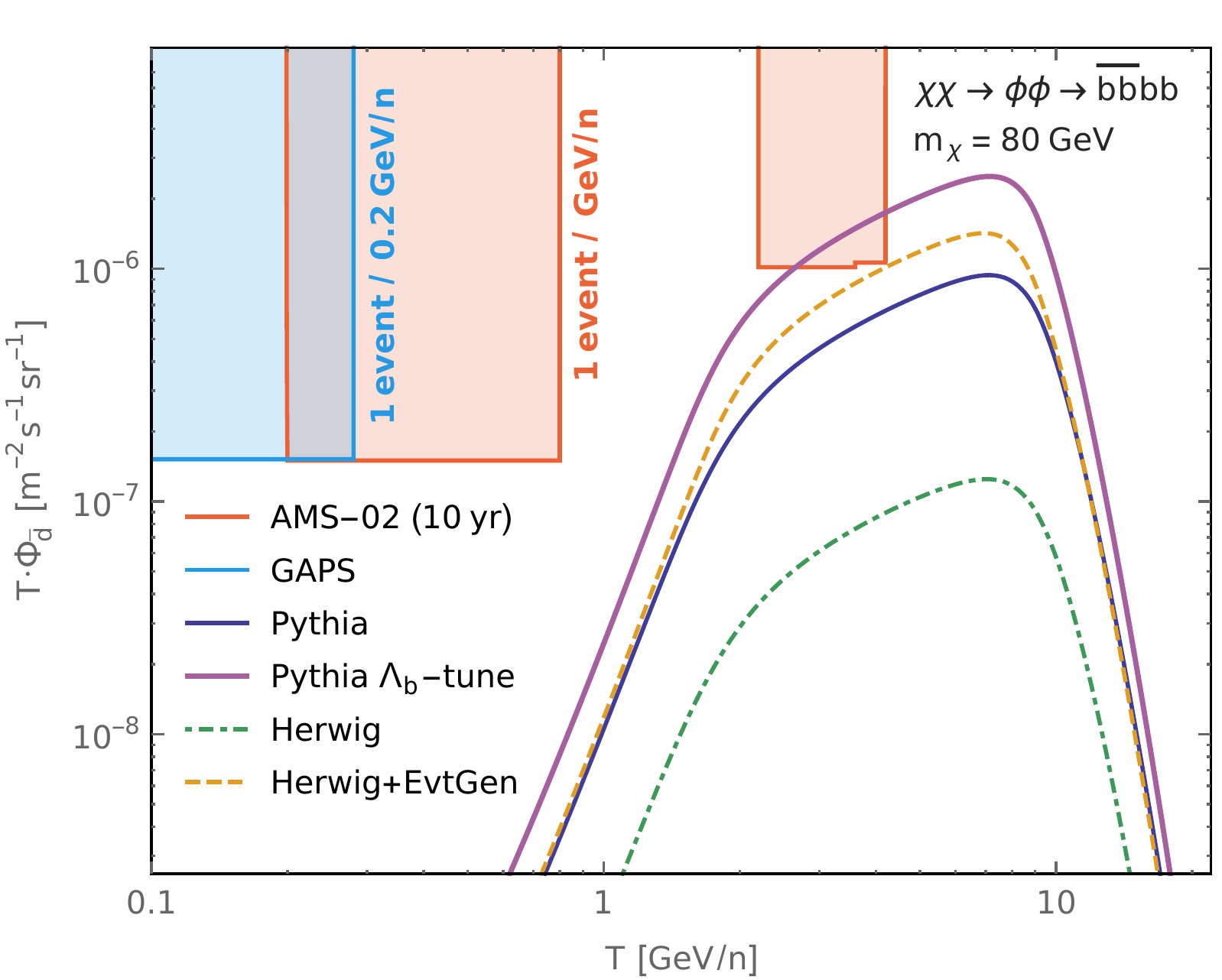}
\vspace{-0.36cm}
\caption{Same as Figure~\ref{fig:fluxes}, but for antideuterons instead of antihelium. The AMS-02 and GAPS antideuteron sensitivities are taken from~\cite{Aramaki:2015laa}.}
\vspace{-0.3cm}
\label{fig:deuterons}
\end{figure*}

Since antihelium carries baryon number $-3$, two diquark-antidiquark pairs must be acquired from the vacuum during hadronization. In the string model such pairs may arise at each factorization step via string breaking, whereas only quark-antiquark pairs are created in the analogous cluster fissioning of Herwig. In the cluster model, new diquarks can emerge in the final decay of clusters into hadrons. However, we find that the resulting cluster mass distribution of $\bar{\Lambda}_b$ decays strongly suppresses multi-baryon final states kinematically. Thus, it appears that the small phase space of antihelium formation via $\bar{\Lambda}_b$ isolates a region in which the differences between baryon production in the string and cluster hadronization models are most stark. We also note that Herwig produces a larger fraction of events that lie above the coalescence threshold. This is linked to a smaller probability of pion-emission associated to the relevant $\bar{\Lambda}_b$ decays compared to Pythia.

While the detailed analysis of the applicability of these models to $\Lambda_b$ decays lies beyond the scope of this work, our preliminary analysis indicates that reasonable changes to the input parameters of the Herwig cluster hadronization model do not qualitatively affect any of the conclusions presented here. Most notably, scans of the {\tt PwtDIquark} parameter (which affects the diquark production probability in cluster decays) and the utilization of the Kupco hadronization model only marginally affect antihelium formation. Thus, we note that our analysis strongly motivates observational and theoretical investigations into the physics of $\Lambda_b$ decays.

Even in the case of the Pythia and Herwig+EvtGen analyses, we note that the predicted antihelium flux still lies below the tentative detection of $\mathcal{O}(10)$ $\hebarthree$ events (and potentially $\mathcal{O}(1)$ $\hebarfour$ events) by AMS-02. However, the significant boosts to antihelium production motivate the careful analysis of $\Lambda_b$ physics. Given the extreme variations in displaced antihelium production between the event generators, it is conceivable that the true antihelium flux even lies above the range obtained in this work. Furthermore, several additional mechanisms previously described in the literature (e.g., Alfv{\'e}nic reacceleration~\cite{Cholis:2020twh}) could potentially be coupled with our results to further enhance the antihelium flux. Finally, we note that our analysis would predict a $\sim$10x larger antihelium event rate if we utilized the more optimistic antihelium sensitivity estimates employed in e.g.~\cite{Cholis:2020twh} (see Appendix~\ref{app:hesensitivity}).

\vspace{0.2cm}

\noindent \emph{Light-Mediator Models --}
One scenario that further increases the antihelium flux are models where dark matter annihilates into light-mediators with masses just above the $\bar{\Lambda}_b$. This significantly increases the fraction of the total dark matter annihilation energy that proceeds through the $\bar{\Lambda}_b$ channel. In Figure~\ref{fig:fluxes} (right), we show the results of such a model, based on the annihilation of 80~GeV dark matter through a mediator of mass 14~GeV into a $\bar{b}\bar{b}bb$ final state. This scenario further boosts the detectable antihelium flux by up to a factor of 3, and further optimizes the $\bar{\Lambda}_b$ production channel while diminishing the effect of prompt antihelium production.

\vspace{0.2cm}
\noindent \emph{Implications for Antideuterons --} The primary goal of our analysis is to examine the effect of $\bar{\Lambda}_b$ baryon decay on antihelium production. This is motivated by the tentative detection of $\mathcal{O}(10)$ antihelium events by AMS-02. However, $\bar{\Lambda}_b$ decays can also produce a significant antideuteron flux. As in the case of antihelium, the boost factor obtained by the $\bar{\Lambda}_b$ is imparted to the antideuterons, significantly enhancing the high-energy antideuteron flux compared to previous predictions.

In Table~\ref{tab:events} and Figure~\ref{fig:deuterons}, we show the integrated and differential antideuteron fluxes produced by our analysis, utilizing the calculation for the antideuteron coalescence momentum discussed in Appendix~\ref{app:coalescence}. We separate our results into contributions from prompt events and $\bar{\Lambda}_b$ baryons. Unlike antiheliums, in the case of antideuterons we also obtain significant contributions from the displaced vertices of $B$ mesons.

\vspace{0.55cm}
\noindent \emph{Discussion --} In this \emph{letter}, we have shown that dark matter annihilations can produce a detectable antihelium flux. Our model assumes no new physics in the dark sector, but instead properly accounts for the contribution of displaced-vertex decays of $\bar{\Lambda}_b$ baryons produced in quark hadronization models. Thus, the results of this study should significantly impact the current assumption that dark matter annihilation events typically produce a negligible anti-nuclei flux.

Moreover, our study significantly alters our understanding of the antinuclei spectrum from dark matter annihilation. Previous analytic and computational studies have focused on prompt antinuclei production~\cite{Donato:1999gy, Carlson:2014ssa, Cirelli:2014qia}, predicting spectra that generically peak at $\lesssim$1~GeV. This spectrum has been utilized to motivate the ``silver-bullet" status of antinuclei searches as a background-free detection strategy, and driven experimental techniques to enhance the low-energy acceptance of balloon-based detectors, such as GAPS~\cite{Aramaki:2015laa}. While our results do not spoil the potential of this strategy, our analysis provides the new opportunity to augment current searches with a brighter, high-energy spectral signature.

While our analysis generically predicts an enhancement of high-energy antihelium, the exact spectrum depends on the dark matter mass and final state. As an example, models of 300~GeV dark matter particle annihilating to $b\bar{b}$ produce a $\bar{\Lambda}_b$-induced antihelium flux which peaks at $\sim$30~GeV/n. Conversely, for lighter dark matter, the antinuclei energy produced by $\bar{\Lambda}_b$ decays is reduced, potentially making the $\bar{\Lambda}_b$ induced antideuteron bump fall within the AMS-02 energy-band.

Finally, while our results are generically produced by leading algorithms such as Pythia and EvtGen, it is notable that standard Herwig analyses produce a negligible (though non-zero) antihelium flux from $\bar{\Lambda}_b$ decay. We have identified the string/cluster hadronization models as the critical difference between these codes, and have found that the small kinematic window for $\bar{\Lambda}_b$ decays to antihelium nuclei isolates a regime where these approaches differ most acutely. Our results motivate a dedicated research program to investigate the decay properties of $\Lambda_b$ baryons and to understand the potential of displaced-vertex antinuclei searches.

\section*{Acknowledgements}

We thank Ilias Cholis for helpful comments involving solar modulation potentials, Philip von Doetinchem for comments regarding the AMS-02 antihelium acceptance and Mirko Boezio, Fiorenza Donato, Carmelo Evoli, Dan Hooper, Alejandro Ibarra,  Michael Korsmeier, Alberto Olivo, Stefano Profumo, Pasquale Serpico, and Andrea Vittino for providing helpful comments on the manuscript. We especially thank Peter Reimitz and Peter Richardson for assistance with the Herwig code.

TL is partially supported by the Swedish Research Council under contract 2019-05135 and the Swedish National Space Agency under contract 117/19. MWW acknowledges support by the Vetenskapsr\r{a}det (Swedish Research Council) through contract No.~638-2013-8993 and the Oskar Klein Centre for Cosmoparticle Physics.

\appendix

\section{Coalescence Momentum}\label{app:coalescence}

In the coalescence model~\cite{Schwarzschild:1963zz}, (anti)nucleons bind if they are produced sufficiently close in phase space. The analytic coalescence model (see e.g.~\cite{Chardonnet:1997dv})
assumes that antinuclei spectra can be determined by the product of single-nucleon spectra (ignoring correlations in multi-nucleon production):
\be\label{eq:analytic}
E_A \frac{d^3 N_A}{dp_A^3} =
B_A \left(E_p\frac{d^3 N_p}{dp_p^3}\right)^{Z}\,\left(E_n\frac{d^3 N_n}{dp_n^3}\right)^{A-Z}
\;,
\ee
evaluated at $\mathbf{p}_{p,n}=\mathbf{p}_A/A$.
The (constant) coalescence factor $B_A$ of the nucleus $A$ accounts for the phase space volume in which coalescence takes place. It can be expressed in terms of the coalescence momentum $p_c$:
\be\label{eq:BA}
B_A= \frac{m_A}{m_p^Z\, m_n^{A-Z}} \left( \frac{\pi}{6} p_c^3\right)^{A-1}\,.
\ee
Given that the single-nucleon spectra are spherically symmetric, the expression for the nucleus spectrum simplifies,
\be
    \frac{dN_A}{dE_A}=\frac{A \,m_A}{m_p^Z\, m_n^{A-Z}} \left(\frac{A \,p_c^3}{24 \,p_A}\right)^{A-1}\left(\frac{dN_{\bar{p}}}{dE_{\bar{p}}}\right)^{Z}\left(\frac{dN_{\bar{n}}}{dE_{\bar{n}}}\right)^{A-Z}\,.
\ee
The analytic coalescence model works reasonably well for hadronic collisions (up to intermediate energies). It is less suited for dark matter annihilation, where final states are emitted in two hard back-to-back jets~\cite{Kadastik:2009ts}. It also fails to describe antinucleus production by the decay of intermediate heavy hadrons.

\begin{figure}[ht!]
\centering
\includegraphics[width=.48\textwidth]{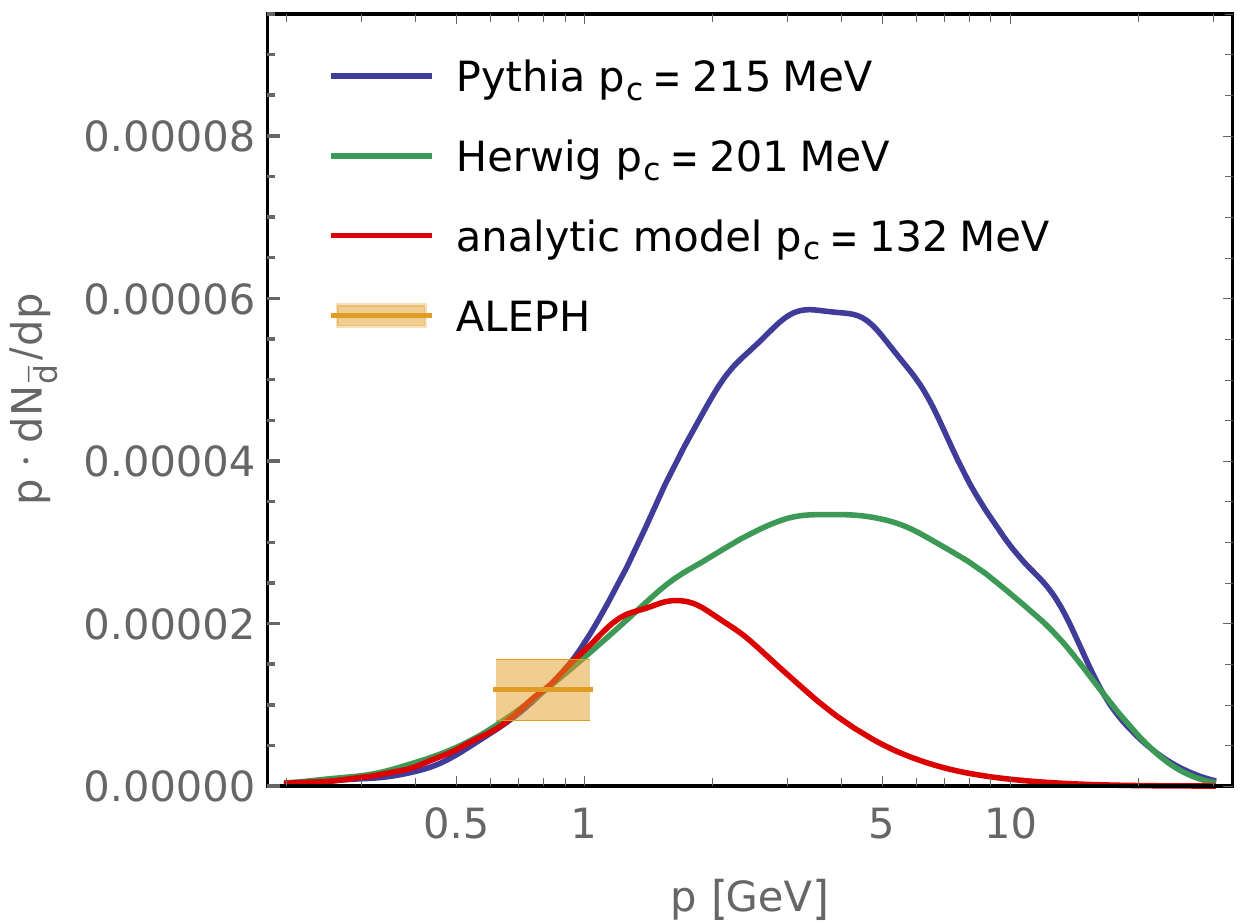}
\caption{Antideuteron spectrum in hadronic $Z$-decays in the event-by-event (Pythia, Herwig) and analytic coalescence models. The coalescence momentum was adjusted to match the ALEPH data.}
\label{fig:aleph}
\end{figure}

\begin{figure}[ht!]
\centering
\includegraphics[width=.48\textwidth]{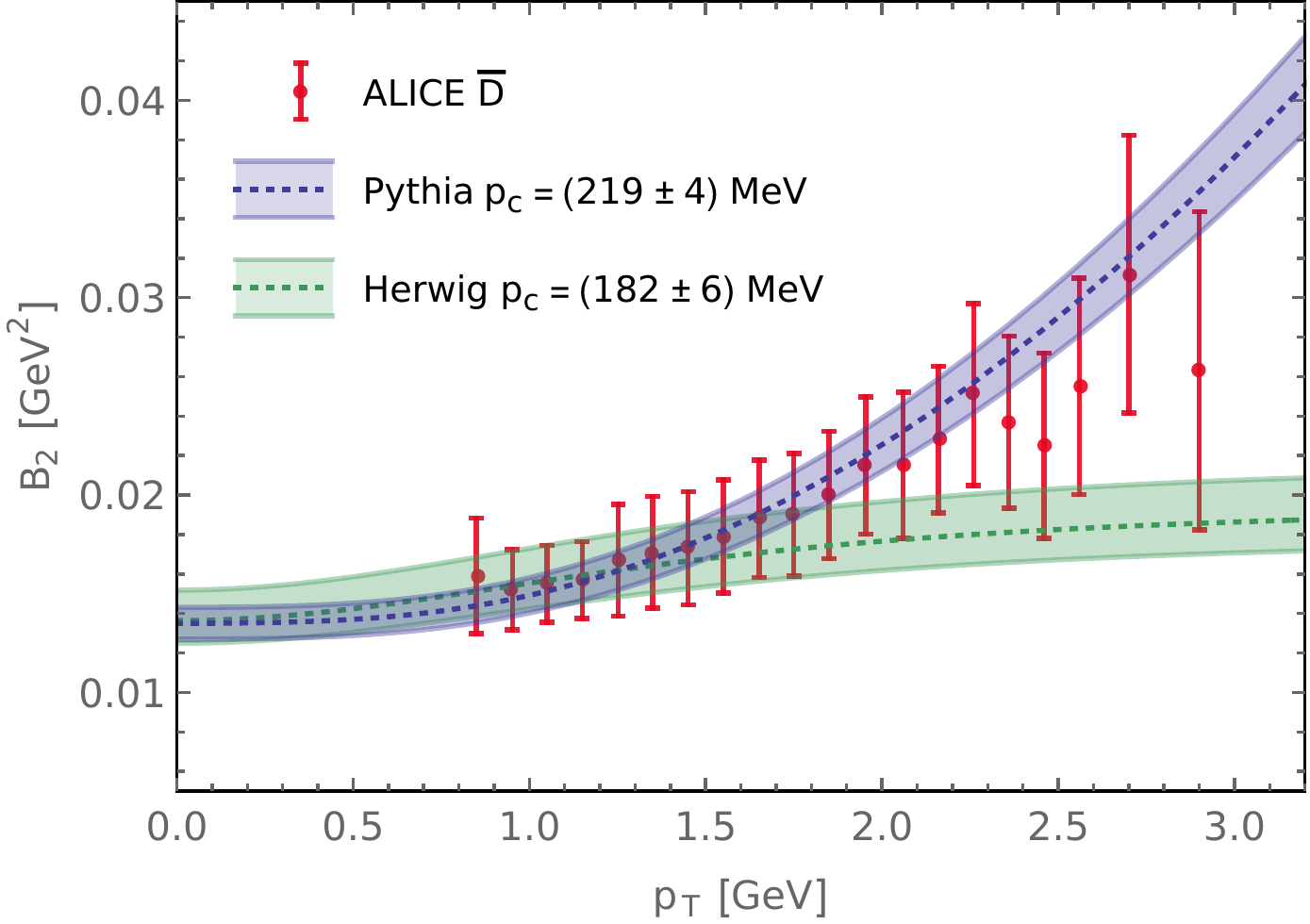}\\[2mm]
\includegraphics[width=.48\textwidth]{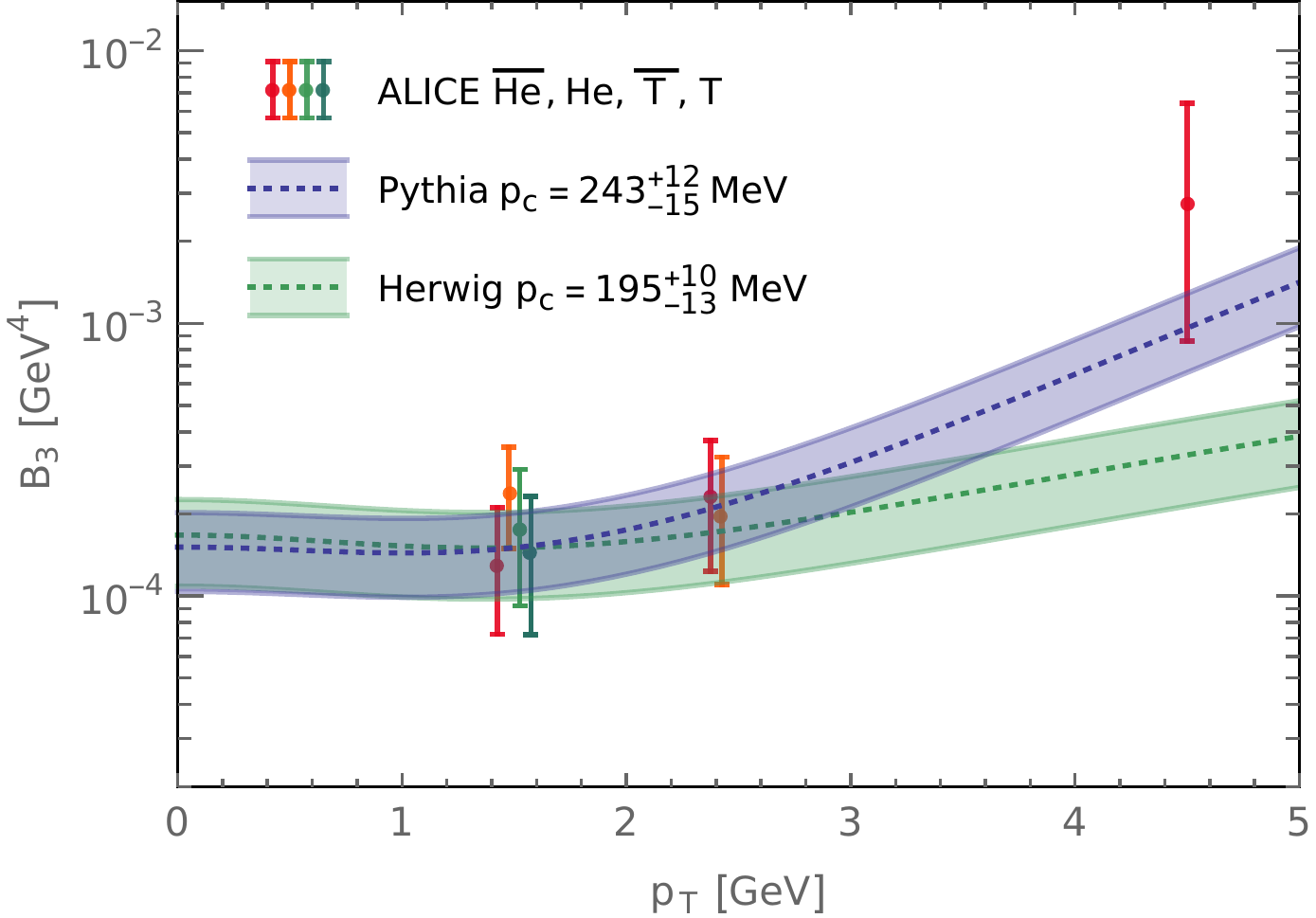}
\caption{Coalescence factor as a function of the transverse momentum for antideuteron (upper panel) and antihelium (lower panel) formation. The ALICE data are shown together with the best fit and uncertainty band in the event-by-event coalescence model (Pythia and Herwig). In the case of $B_3$, (anti)triton and (anti)helium data were combined to increase the statistics (for better visibility, data taken at the same transverse momentum appear slightly shifted in the figure).}
\label{fig:alice}
\end{figure}

In this work, we employ the event-by-event coalescence model. Dark matter annihilation is simulated with the Monte Carlo generators Pythia and Herwig. For Pythia we also consider the $\Lambda_b$-tune which was calibrated to reproduce the $\Lambda_b$ production rate at LEP (see main text). In addition we include a Herwig implementation in which $\Lambda_b$ decays are handled by the specialized tool EvtGen. The event-by-event coalescence model has the advantage that it fully incorporates correlations in the multi-nucleon spectra as provided by the event generator. The following coalescence conditions are imposed on an event-by-event basis.
\begin{itemize}
  \item Nucleons have to stem from the same interaction vertex. This can either be the initial vertex or the decay vertex of the same intermediate particle (e.g.\ a $\bar{\Lambda}_b$).
  \item A multi-nucleon system forms a bound state if -- in its center-of-mass frame -- the momentum of each nucleon is smaller than $p_c/2$ (deuteron) or $2^{1/6}p_c/2$ (helium-3, triton).
\end{itemize}
We note that our coalescence condition for helium differs by $2^{1/6}$ compared to~\cite{Carlson:2014ssa}. Our definition appears preferable since it yields a direct correspondence between the analytic and the event-by-event coalescence models. The factor $2^{1/6}$ for helium ensures that the coalescence volume matches a multi-dimensional sphere of radius $p_c/2$. We explicitly verified that with the coalescence condition defined above, we can reproduce the predictions of the analytic coalescence model once we erase correlations through event-mixing.\footnote{Event mixing means that we randomly redistribute nucleons between generated events before applying the coalescence condition.}

The coalescence momentum is a free parameter and must be tuned to experimental data. From the viewpoint of antinuclei formation, dark matter annihilation bears resemblance to electron-positron collision. ALEPH has measured $e^+ e^-\rightarrow \dbar$ at the $Z$-resonance~\cite{Schael:2006fd}. The antideuteron production rate per hadronic $Z$-decay in the momentum-range $p_{\dbar}=0.62-1.03\gev$ and polar angle interval $|\cos\theta|<0.95$ was determined as:
\be
R_{\dbar}=(5.9\pm 1.9)\times 10^{-6}\,.
\ee

From the ALEPH data we determine the antideuteron coalescence momentum for dark matter ($\chi$) annihilation (following a similar approach as~\cite{Kadastik:2009ts,Dal:2012my,Fornengo:2013osa}):
\begin{align}\label{eq:pcaleph}
p_c \left(\chi\chi\rightarrow\dbar \right) &= 215^{+19}_{-23} \mev \quad \text{(Pythia)}\nonumber\\
p_c \left(\chi\chi\rightarrow\dbar \right) &= 124^{+11}_{-13} \mev \quad \text{(Pythia $\Lambda_b$-tune)}\nonumber\\
p_c \left(\chi\chi\rightarrow\dbar \right) &= 201^{+20}_{-24} \mev \quad \text{(Herwig)}
\end{align}
In the Pythia $\Lambda_b$-tune model, the increased diquark formation probability scales up the deuteron spectrum which is compensated by the significantly reduced $p_c$. The Herwig result, on the other hand, does not depend on whether the EvtGen option is enabled. We note that the deuteron spectra obtained with Pythia and Herwig look significantly different even when normalized to the ALEPH measurement. This is illustrated in Figure~\ref{fig:aleph}. For comparison we also included the spectrum of the analytic coalescence model. The spectra corresponding to the Pythia and Herwig modifications are not shown separately since they exhibit a similar shape as the respective default configurations.

Antihelium production has so far only been measured in $pp$ collisions by ALICE~\cite{Acharya:2017fvb} at a high center-of-mass energy of $\sqrt{s}=7\:\text{TeV}$. Due to the major differences in the initial state compared to standard dark matter annihilation models, we refrain from directly normalizing dark matter induced antihelium fluxes to the ALICE result. However, the ratio of deuteron and helium coalescence momenta should only weakly depend on the initial state. Therefore, we extract the antihelium coalescence relevant for dark matter annihilation (simply denoted by $p_c$ in the main part) from the following expression:
  \be\label{eq:pchelium}
  p_{c}= p_c \left(\chi\chi\rightarrow\dbar \right)\times \frac{p_c\left(pp\rightarrow\hebar\right)}{p_c\left(pp\rightarrow\dbar\right)}\,.
  \ee
ALICE determined the coalescence factor $B_A$ ($A=2,3$) defined as in eq.~\eqref{eq:analytic}. In the analytic coalescence model $B_A$ is a constant -- in conflict with the ALICE data in which it exhibits a significant transverse-momentum-dependence. The latter can be explained by correlations in the antinucleon production which are accounted for in the event-by-event coalescence model. Generalizing~\eqref{eq:BA}, we define
\be
B_A= \xi(\mathbf{p}_A)\,\frac{m_A}{m_p^Z\, m_n^{A-Z}} \left( \frac{\pi}{6} p_c^3\right)^{A-1}\,,
\ee
where we now include an additional momentum-dependent correlation factor $\xi$. We determine $\xi$ with the event generators Pythia and Herwig and then fit $p_c$ to the ALICE data (see Figure~\ref{fig:alice}).

The coalescence momenta extracted from our fit are thus:
\begin{align}
p_{c}\left(pp\rightarrow \dbar\right) &= (219\pm 4)\mev\quad \text{(Pythia)} \nonumber\\
p_{c}\left(pp\rightarrow \dbar\right) &= (182\pm 6)\mev\quad \text{(Herwig)} \nonumber\\
p_{c}\left(pp\rightarrow \hebar\right) &= 243^{+12}_{-15}\mev\quad \text{(Pythia)}\nonumber \\
p_{c}\left(pp\rightarrow \hebar\right) &= 195^{+10}_{-13}\mev\quad \text{(Herwig)}\nonumber \\
\end{align}
We verified that $p_c(pp\rightarrow\hebar)/p_c(pp\rightarrow\dbar)$ agrees between default Pythia and the $\Lambda_b$-tune. Similarly, the ratio in Herwig is not affected if the EvtGen option is enabled. The antihelium coalescence momentum for dark matter annihilation as derived from~\eqref{eq:pchelium} is stated in~\eqref{eq:pcpythiaherwig}.

\newpage
\section{Helium Sensitivity}\label{app:hesensitivity}
\vspace{-0.2cm}
\noindent We define the antihelium acceptance $\eta_{\hebar}$ such that
\vspace{-0.2cm}
\be
 N_{\hebar} = \int d\mathcal{R}\: \Phi_{\hebar}\; \eta_{\hebar}(\mathcal{R})\,,
\ee
where $N_{\hebar}$ denotes the observed number of antihelium events, $\Phi_{\hebar}$ the antihelium flux and $\mathcal{R}$ the rigidity. The AMS-02 projected 10-year constraint, translated from the expected 18~year limit of $5\times 10^{-10}$~\cite{Kounine2011} is:
\vspace{-0.2cm}
\be
\left(\frac{\hebar}{\text{He}}\right)_{\text{max}}=9\times 10^{-10}
\ee
\vspace{-0.2cm}

at 95$\%$ CL (corresponding to 3 $\hebar$ events) in the rigidity range $\mathcal{R}=1-50\:\text{GV}$. Assuming that antihelium and helium acceptances have a similar rigidity-dependence (i.e.\ $\eta_{\hebar}(\mathcal{R})/\eta_{\text{He}}(\mathcal{R})=\text{const}$) we derive:
\be
\eta_{\hebar}(\mathcal{R})=\frac{3}{(\hebar/\text{He})_{\text{max}}}\; \frac{\eta_{\text{He}}(\mathcal{R})}{N_{\text{He}}}\,.
\ee
Extracting the number of observed helium events $N_{\text{He}}$ and the helium acceptance from the supplemental material of~\cite{Aguilar:2015ctt}, we determine the antihelium acceptance as a function of rigidity. The sensitivity shown in Figure~\ref{fig:fluxes} is obtained after converting rigidity to kinetic energy per nucleon.

\bibliography{main}

\end{document}